\documentclass[a4paper, 11pt]{article}

\pdfoutput=1

\usepackage{amsmath, amssymb, bbm, setspace, geometry, amsfonts, enumerate, amsthm, subfigure, caption, rotating, algorithm, changebar, lipsum, multicol, ifpdf, natbib}

\newtheorem{theorem}{Theorem}[section]
\newtheorem{lemma}[theorem]{Lemma}
\newtheorem{proposition}[theorem]{Proposition}

\theoremstyle{definition}
\newtheorem{definition}[theorem]{Definition}

\theoremstyle{remark}

\renewcommand{\qed}{\hfill \ensuremath{\Box}}

\newcommand{\supp}{\text{supp}}

\newcommand{\E}{\mathbb{E}}

\newcommand{\var}{\text{Var}}

\newcommand{\Z}{\mathbb{Z}}

\newcommand{\N}{\mathbb{N}}

\newcommand{\norm}{\mathcal{N}}

\newcommand{\Or}{\mathcal{O}}

\newcommand{\leftb}{\!\left}

\newcommand{\cI}{\tilde{c}}

\newcommand{\HF}{\mathcal{F}}

\newcommand{\HFI}{\mathcal{H}}

\newcommand{\HFB}{\mathcal{B}}

\newcommand{\diff}{\text{d}}

\newcommand{\half}{{}^1 \! /_{\!2}}

\newcommand{\third}{{}^1 \! /_{\!3}}

\newcommand{\kT}{{}^k \! /_{\!T}}

\begin{document}

\title{Bayesian Wavelet Shrinkage of the Haar-Fisz Transformed Wavelet Periodogram}
\author{Guy P. Nason and Kara N. Stevens}

\maketitle

\section*{Abstract}

It is increasingly being realised that many real world time series are not stationary and exhibit evolving second-order autocovariance or
spectral structure. This article introduces a Bayesian approach for modelling the evolving wavelet spectrum of a locally stationary
wavelet time series. Our new method works by combining the advantages of a Haar-Fisz transformed spectrum with a simple, but powerful, Bayesian
wavelet shrinkage method. Our new method produces excellent and stable spectral estimates and this is demonstrated via simulated data and
on differenced infant ECG data. A major additional benefit of the Bayesian paradigm is that we obtain rigorous and useful credible intervals of the evolving spectral structure. We show how the Bayesian credible intervals provide extra insight into the infant ECG data.

\section{Introduction}

For a real-life time series it is sometimes difficult to determine whether the underlying process is really stationary using only observations from a section of the process. Often, the spectral behaviour of a real-life time series can change from one time point to another and nonstationarity may only become apparent with continued observation. If we disregard the stationarity assumption, there are an abundance of different models that can be considered. One class of nonstationary models, which we consider here, are the locally stationary processes with slowly evolving second-order structure. Two prominent sub-classes are the locally stationary (Fourier) processes due to \citet{Dahlhaus:97} and the locally stationary wavelet (LSW) processes due to \citet{Nason:00}. However, nonstationary Fourier processes have a long history see, e.g. \citet{Page:52, Silverman:57, Priestley:65}. Reviews can be found in \citet{Nason:99} and \citet{Dahlhaus:12}. The second-order structure of a time series can be assessed via the (auto-)covariance or spectrum, and accurate specification and estimation of these quantities is particularly important to improve our understanding of the data.

This article assumes that a time series can be modelled by a LSW process and considers the estimation of the associated evolutionary wavelet spectrum (EWS). As is the case for stationary spectral estimation obvious `raw' estimators are not statistically consistent and require smoothing. For example, \citet{Nason:00} introduced a kind of `method of moments' spectral estimator and used wavelet shrinkage to smooth it and \citet{Fryzlewicz:03} used kernel smoothing to produce estimates for forecasting. See also \citet{Bellegem:08} who introduce a pointwise estimator. \citet{Fryzlewicz:06} introduced a powerful new approach based on Haar-Fisz transformation of the raw wavelet periodogram and essentially using universal thresholding \citep{DonohoJohnstone:94a} on the Haar-Fisz coefficients.

This article builds on the \cite{Fryzlewicz:06} work by using Bayesian wavelet shrinkage to bear on the Haar-Fisz coefficients and does so for two reasons. First, recent Bayesian wavelet shrinkage techniques based on the Berger-M\"{u}ller prior and empirical marginal maximum likelihood determination, such as \citet{Johnstone:05}, show dramatic performance improvements over earlier concepts such as universal thresholding. The Bayesian approach uses priors well-adapted to the known  mathematical \textit{theory} underlying wavelet coefficients of a wide class of functions from Besov scales. Secondly, the coherent Bayesian approach permits rational and effective quantification of  credible intervals for the EWS. Our simulation results and results on real data show good performance and new insights.

Section~\ref{sec:LSW.p} reviews the locally stationary wavelet model and the associated evolutionary wavelet spectrum and the wavelet periodogram. Section~\ref{sec:HF} briefly reviews the Haar-Fisz transformation at establishes notation for subsequent Bayesian wavelet shrinkage. Section~\ref{sec:BWS} first reviews wavelet shrinkage and Bayesian wavelet shrinkage and then describes each of the components of our Bayesian wavelet shrinkage method adapted for the Haar-Fisz-transformed spectral coefficients. Section~\ref{sec:Sim} outlines some implementation issues, presents a simulation and analyses an infant ECG data set and compares it to earlier analyses. Finally, section~\ref{sec:conc} concludes and provides some ideas for further developments.

\section{Locally Stationary Wavelet Processes}
\label{sec:LSW.p}

Locally stationary wavelet (LSW) processes were introduced by \citet{Nason:00}, and extended to encompass a larger range of processes in \citet{Bellegem:08}
which we use here. As in \citet{Nason:00} assume that the wavelets used are \cite{Daubechies} compactly supported, and that the length of the support for any wavelet $\psi_{j,0}$ is equal to $\mathcal{L}_j := | \supp (\psi_{j,0}) |$. Therefore, if we have $J$ scales, where $1$ is the finest scale, then $|\supp (\psi_{j,k}) | = \mathcal{L}_j = (2^j - 1)(\mathcal{L}_1- 1) + 1 \ \forall \, j \geq 0$, where $\mathcal{L}_1$ is the support at the finest scale. Here $\N$ is the set of natural numbers $\{1, 2, 3, \ldots\}$.

\begin{definition}[The Locally Stationary Wavelet Process]
	 A LSW process is a sequence of doubly indexed stochastic processes, $\{ X_{t,T}\}_{t=0, \ldots,T-1}$, where $T = 2^J$ for some $J \in \N$. This process has the representation
	\begin{equation}
		X_{t,T} \ = \ \sum_{j=1}^{\infty} \sum_{k = -\infty}^{\infty} w_{j,k:T} \ \psi_{j,k-t}^{(s)} \ \xi_{j,k},
		\label{eq:LSW.process}
	\end{equation}
	where $\psi_{j,k-t}^{(s)}$ is a discrete non-decimated family of wavelets for scale $j \in \N$, location $k \in \Z$ based on a mother wavelet, $\psi(t)$, of compact support, which we shall refer to as the \textit{synthesis wavelet}; and $\xi_{j,k}$ is a Gaussian random zero mean orthonormal increments sequence. The component $w_{j,k:T} \, \xi_{j,k} $ can be thought of as a random amplitude of the oscillation $\psi_{j,k-t}^{(s)}$. 

	The quantities in equation \eqref{eq:LSW.process} possesses the following properties:
	\begin{enumerate}[(a)]
		\item $\E[\xi_{j,k}]=0, \ \forall \, j\in \N, \ k \in \Z \ (\Rightarrow \ \E[X_t]=0)$.
		\item $\E[\xi_{j,k}, \xi_{j',k'}]= \delta_{j,j'} \, \delta_{k,k'}, \ \forall \, j,j' \in \N, \ k, k' \in \Z$.
		\item For each $j \in \N$ there exists a function $W_j(z)$ for $z \in \left(0,1\right)$, that possesses the following properties
			\begin{enumerate}[i.]
				\item 
					\[
						\sum_{j=1}^{\infty} |W_j(z)|^2 < \bar{C} \quad \text{uniformly in } z \in \left(0,1\right).
					\]
				\item There exists a sequence of constants $C_j$ such that for each $T$
					\[
						\sup_k | w_{j,k;T} - W_j (\kT) | \leq \frac{C_j}{T}.
					\]
				\item The total variation (TV) of $W_j^2(z)$ is bounded by a constant $L_j$, that is
					\begin{align*}
						TV(W_j^2) & := \sup \leftb\{\sum_{i=1}^I |W_j^2(a_i)-W_j^2(a_{i-1})|:0 <a_0<\ldots<a_I <1, I \in \mathbb{N}\right\} \\
						& \leq \ L_j.
					\end{align*}
				\item  The constants $C_j$ and $L_j$ satisfy
					\[
						\sum_{j=1}^{\infty} \mathcal{L}_j (\mathcal{L}_j L_j + C_j) \leq \rho < \infty.
					\]
			\end{enumerate}
		\end{enumerate}
	\label{def:LSW.processes}
\end{definition}
The time evolution of LSW processes is governed by the time-scale varying evolutionary wavelet spectrum which we define next.

\subsection{Evolutionary Wavelet Spectrum and its Estimation}
\label{EWS}

The evolutionary wavelet spectrum (EWS) measures the `contribution to the variance' of $X_{t,T}$ at scale level $j\in \N$ and location $z \in (0,1)$ and is defined as follows.
\begin{definition}[Evolutionary Wavelet Spectrum]
	The EWS is defined by
	\begin{equation}
		S_j(z) \ = \left|W_j(z)\right|^2 \quad \forall  \, j \in \N \text{ and } z \in \left(0,1\right).
	\end{equation}
	\label{def:EWS}
\end{definition}

Estimation of the EWS can be achieved by first computing the raw wavelet periodogram, defined as follows.
\begin{definition}[Raw Wavelet Periodogram]
	The raw wavelet periodogram is defined as
	\begin{equation}
		I_{j,k;T} \  = \left| \sum_{t=-\infty}^{\infty}X_{t,T} \, \psi_{j,k-t}^{(a)} \right|^2,
		\label{eq:rwp}
	\end{equation}
	where $X_{t,T}=0$ for $t \neq 0, \ldots, T-1, \ j = 1, \ldots, J, \ k = 0, \ldots, T-1, \ J = \log_2(T)$ and $\psi_{j,k}^{(a)}$ is a discrete non-decimated family of wavelets we shall refer to as the \textit{analysis wavelet}.
	\label{def:I}
\end{definition}

In theory, the analysis wavelet from \eqref{eq:rwp} is the same as the synthesis wavelet in \eqref{eq:LSW.process} . However, often in practice the synthesis wavelet is unknown. For the purposes of our analysis we shall assume the synthesis wavelet is known and equivalent to the analysis wavelet. The raw wavelet periodogram, $I_{j,k}$, is a biased estimator of the EWS, but can be made asymptotically unbiased after simple correction which we will explain next. To proceed with this, the autocorrelation wavelet (ACW) is defined as follows.
\begin{definition}[Discrete Autocorrelation Wavelet] 
	The ACW at scale $j \in \N$ at lag $\tau \in \Z$ is defined by
	\[
		\Psi_j(\tau) \ = \ \sum_{k=-\infty}^{\infty}{\psi_{j,k} \psi_{j,k-\tau}}.
	\]
	\label{def:dACW}
\end{definition}
The discrete ACW determines the autocorrelation of a wavelet at a particular scale, $j$ and different lags, $\tau$. The discrete ACW provides a family of symmetric, compactly supported, positive semi-definite functions on $\tau \in \Z$. Further theoretical details  can be found in \citet{Nason:00} and
\citet{Eckley:05}. To form an asymptotically unbiased estimator of the spectrum we require the inner product matrix of the ACW defined as
follows.
\begin{definition}[The Inner Product Matrix] 
	The operator $A= (A_{j,l})_{j,l \geq 0}$ is defined by
	\begin{equation}
		A_{j,l} \ = \left\langle \Psi_j, \Psi_l \right\rangle = \ \sum_\tau \Psi_j(\tau) \; \Psi_l(\tau).
		\label{eq:A}
	\end{equation}
	and the $J$-dimensional matrix is $A_J = (A_{j,l})_{j, \; l = 1, \ldots, J}$.
	\label{def:A}
\end{definition}

Then using definitions \ref{def:LSW.processes} and \ref{def:A}, proposition 3.3 of \citet{Nason:00} shows that
\begin{equation}
	\E\left[I_{j,k}\right] = \ \sum_l A_{j,l} \, S_l (z) + \mathcal{O} \left(T^{-1}\right), \quad \forall \; z \; \in \; [0,1),
	\label{eq:E.I}
\end{equation}
for $j\in \N, k\in\Z$, where $A$ is calculated using the chosen analysis wavelet and the variance of the wavelet periodogram is given by
\begin{equation}
	\text{var}\left[I_{j,k;T}\right] = \ 2 \left\{ \sum_l A_{j,l} \, S_l(z) \right\}^2 + \Or\leftb(\frac{2^j}{T}\right), \qquad j \geq 0
	\label{eq:V.I}
\end{equation}
This result implies that as the sample size increases ($T \rightarrow \infty$) the variance does not vanish. \cite{Nason:00} show that the obvious asymptotically unbiased estimator $A_J^{-1} \mathbf{I}_k$ for $\{ S_j (\kT) \}$ where $\mathbf{I}_k = (I_{1, k}, \ldots, I_{J, k})$ is not statistically
consistent. As is typical in spectral analysis in time series the periodogram needs to be smoothed to obtain consistency.

\subsection{Wavelet Periodogram Smoothing}

Various techniques have already been developed to smooth the wavelet periodogram, such as those by \citet{Nason:00, Fryzlewicz:06, Bellegem:08}.
\citet{Bellegem:08} is theoretically attractive but tricky to implement in practice.

In \citet{Nason:00} each level, $j$, of the raw wavelet periodogram is smoothed as a function of $z$ using translation-invariant (TI) de-noising \citep{Coifman1}. Non-linear wavelet shrinkage is performed on the approximately $\chi_1^2$ distributed raw wavelet periodogram then bias corrected by the inner product matrix $(A^{-1})$.  An appropriate threshold for the shrinkage was determined in \citet[Theorem 3.4]{Nason:00}. The technique raises a number of questions, such as what is an appropriate wavelet? \citet{Nason:00} believe that smoother wavelets, such as Daubechies extremal phase with 10 vanishing moments, help to avoid `leakage' of power into the surrounding scales because of their short support in the Fourier domain. They also produce less spiky and variable estimates in their example. 

\citet{Fryzlewicz:06} suggested applying the soft shrinkage rule upon the Haar-Fisz coefficients of the raw wavelet periodogram, using a scale dependent threshold. The methodology produced an estimator which was mean-square consistent, rapidly computable, easy to implement and performs well in practice. However, the theoretical validation of this technique was restricted to locally stationary processes with a time-varying, but piecewise constant form.

The Haar-Fisz transform in \citet{Fryzlewicz:06} is very attractive producing transformed periodogram ordinates that are very close to being uncorrelated
and Gaussian. We apply Bayesian wavelet shrinkage to this enticing situation and not having to worry about first order estimation error in the variance.

\section{Spectral Normalisation using the Haar-Fisz Transform}
\label{sec:HF}

The Haar-Fisz transformation works by normalising the wavelet coefficients of a signal to obtain elements that are close to Gaussian and have near-constant variance. We adapt the definition from \cite{Fryzlewicz:06} which applies the Haar-Fisz transform to the raw wavelet periodogram $I_{j,k}$ as follows.

\begin{enumerate}
	\item Let $c_{J,m}:= I_{j,k}$ for $m = 0, \ldots, T-1$, where $T = 2^J$
	\item For $l = (J-1), \ldots, 0$, recursively for the vectors
		\begin{align*}
			& d_{l,m} = \frac{c_{l+1,2m} - c_{l+1,2m+1}}{\sqrt{2}}, \\
			& c_{l,m} = \frac{c_{l+1,2m} + c_{l+1,2m+1}}{\sqrt{2}},
		\end{align*}
		where $m = 1, \ldots, 2^l-1$, and $d_{l,m}$ and $c_{l,m}$ are the Haar wavelet and scaling coefficient of the raw wavelet periodogram at scale $j$, respectively.
	\item Divide the wavelet coefficients by the scaling coefficients to produce the Haar-Fisz coefficients
		\begin{equation}
			f_{l,m} = \frac{d_{l,m}}{c_{l,m}},
		\end{equation}
		for $c_{l, m} \neq 0$. For $c_{l, m} = 0$ set $f_{l, m} = 0$.
	\item For $l = 0, \ldots, J-1$, recursively form the vectors $\cI_{l-1}$:
		\begin{align*}
			c_{l+1, 2m} & = c_{l,m} + f_{l,m} \\
			c_{l+1, 2m-1} & = c_{l,m} - f_{l,m}
		\end{align*}
		where $c_{0,0} = c_{0,0}$ and
		where $m = 1, \ldots, 2^l$,
		\item Define $\HFI_{j,k} = c_{J,m}, \ m = 1, \ldots, 2^J$.
\end{enumerate}
Let $\HF$ denote the non-linear invertible Haar-Fisz operator, hence $\HFI_{j,k} = \HF I_{j,k}$.

\cite{Fryzlewicz:06} model the raw wavelet periodogram  as
\[
	I_{j,k} \ \approx \ R_{j}(z) \, Z_{j,k}^2,
\]
where $R_{j}(z) = (AS)_j(z), \ z= \kT$ and $Z_{j,k}^2 \sim \chi_1^2$,
for $j \in \N, \ k =1, \ldots, 2^J=T$.

Proposition 6.1 in \citet{Fryzlewicz:06} details a number of properties possessed by $\HFI$. Property 6.1(2) states the Haar-Fisz transformation possesses the log-like property, which suggests the a potential model for the $\HFI$ is
\begin{equation}
	\HFI_{j,k} = \HFB_{j}(z) + e_{j,k}
	\label{eq:FHF.I}
\end{equation}
for $j = 1, \ldots, J$ and $k = 1, \ldots, 2^J$, where $\HFB_{j}(z)=\HF R_{j}(z), \ z= \kT$ and $e_{j,k} = \HF Z^2_{j,k}$. As the distribution of $\HFI_{j,k}$ is approximately $\norm(\HFB_{j,k}, \sigma^2_j)$, $e_{j,k}$ are approximately uncorrelated with $e_{j,k} \approx \norm(0, \sigma^2_j)$, due to Proposition 6.1 (3,4,5) from \cite{Fryzlewicz:06}. Model~\eqref{eq:FHF.I} is conducive to Bayesian wavelet shrinkage as explained next.

\section{Bayesian Wavelet Shrinkage}
\label{sec:WS}

\subsection{Brief Review of Wavelet Shrinkage}

Wavelet shrinkage is a form of nonparametric regression introduced in a series of seminal articles such as \cite{DonohoJohnstone:94a,DonohoJohnstone:95}.
See \cite{Vidakovic} or \cite{Nason2} for more details and further references. Suppose we have a set of noisy observations, $\mathbf{y} = (y_1, \ldots , y_n)$ of an unknown function $f(x)$, taken at regularly spaced locations, denoted by $\mathbf{x} = (x_1, \ldots, x_n)$. In our context, we can use the well-known additive signal-plus-noise model for \textit{each} scale-level, $j$, in~\eqref{eq:FHF.I}:
\[
	y_i = f(x_i) + e_i \qquad \qquad \text{for} \ i = 1, \ldots, n,
\]
where $\mathbf{e} = (e_1, \ldots, e_n)$ are random variables which are usually assumed to be iid with zero mean and  some variance $\sigma^2$. The aim is to devise an estimator $\hat{f}(x)$ to recover the signal $f$ (also known as $\HFB$) from the noisy observations $y_i$ ($\HFI$). Wavelet shrinkage is very simple and the estimator can be obtained by the following three steps.
\begin{enumerate}
	\item Apply the discrete wavelet transformation (DWT) to noisy data $\mathbf{y}$, giving
		\[
			\mathbf{d} = \boldsymbol{\beta} + \boldsymbol{\varepsilon},
		\]
		where $\mathbf{d} = W\mathbf{y}$, $\boldsymbol{\beta} = Wf(\mathbf{x})$, $\boldsymbol{\varepsilon} = W\mathbf{e}$ and $W$ is the orthogonal DWT matrix for a particular \textit{smoothing wavelet} (SW). The vector $\mathbf{\beta}$ are considered to be the `true' wavelet coefficients, $\mathbf{d}$ are the noisy empirical wavelet coefficients.
	\item Apply a shrinkage method and threshold (such as hard shrinkage and the universal threshold) to the noisy coefficients, $\mathbf{d}$, to obtain estimates, $\hat{\boldsymbol{\beta}}$, of the wavelet coefficients $\boldsymbol{\beta}$.
	\item Apply the inverse DWT to the estimated coefficients $\hat{\boldsymbol{\beta}}$ to obtain an estimate, $\hat{f}(x)$, of the underlying function $f(x)$ at the data points $\mathbf{x}$.
\end{enumerate}
To enable us to obtain good estimates with a sound basis for obtaining credible intervals we adopt a \textit{Bayesian} wavelet shrinkage approach as described next.

\subsection{Bayesian Wavelet Shrinkage}
\label{sec:BWS}

Bayesian statistical methods start with existing \textit{prior} knowledge of model parameters ($\boldsymbol{\beta}$), which are updated using the data ($\mathbf{y}$) to give \textit{posterior} knowledge. The resulting posterior knowledge can be used to  interpret  these parameters. The model commonly used for Bayesian inference is
\begin{equation}
	p (\boldsymbol{\beta} | \mathbf{y} ) = \frac{p(\mathbf{y} | \boldsymbol{\beta}) p (\boldsymbol{\beta})}{\int_Y p(\mathbf{y} | \boldsymbol{\beta} ) p (\boldsymbol{\beta}) \, d\mathbf{y}},
\label{eq:genbayes}
\end{equation}
where $p(\mathbf{y} | \boldsymbol{\beta} )$ is the \textit{likelihood}, $p(\boldsymbol{\beta})$ is the \textit{prior density} function and $p(\boldsymbol{\beta} | \mathbf{y})$ is the \textit{posterior density} function of $\boldsymbol{\beta}$ given $\mathbf{y}$. Confidence, or more properly credible, intervals can be obtained from the upper and lower tail quantiles of the posterior distribution.

Adopting a Bayesian approach for wavelet shrinkage has become increasingly popular for wavelet denoising due to its excellent theoretical and practical properties, see \cite{Chipman:97}, \cite{Vidakovic98}, \cite{Clyde:99}, \cite{Muller:99}, \cite{Ruggeri} and \cite{Johnstone:05}, for example. Bayesian wavelet shrinkage has also been utilized for stationary spectral estimation in \cite{Pensky:07} and for credible intervals for regression by \cite{Barber:01}, \cite{Semadeni:04} and \cite{Davison:09}. The usual procedure is to place a prior distribution on the wavelet coefficients, use the Bayesian paradigm specified by~\eqref{eq:genbayes} with the necessary components specified as follows to enable us to derive a closed-form expression for the posterior means and variance. For parts of our specification below we shall utiliize the empirical Bayes approach from \cite{Johnstone:05}.

\subsection{Regression Model}

We shall apply Bayesian wavelet shrinkage to the Haar-Fisz transformed wavelet periodogram, $\HFI$. Taking the DWT of \eqref{eq:FHF.I}, for a particular scale $j$, we obtain
\begin{equation}
	h_{l,m} = \beta_{l,m} + \varepsilon_{l,m},
	\label{eq:DWT.HFI}
\end{equation}
where $h_{l,m} = (W\HFI_j)_{l,m}$, $\beta_{l,m}=(W\HFB_j)_{l,m}$, $\varepsilon_{l,m} = (We_j)_{l,m}$ for scales $l = 0, \ldots, J-1$ and locations $m = 1, \ldots, 2^l$, and $W$ is the $T \times T$ orthogonal DWT matrix associated with some \cite{Daubechies} compactly supported wavelet. Due to the orthogonality of the wavelet transformation and the approximate error structure of the $e_{j,k}$ noted above, the distribution of the wavelet-transformed error is approximately $\varepsilon_{l,m} \sim \norm(0, \nu_l^2)$, where $\nu_l^2 = 2^{J-l}\sigma_j^2$. For notational clarity we shall cease mention of the scale index $j$. However, it should be remembered we are applying Bayesian wavelet shrinkage scale-by-scale $j$ to~\eqref{eq:FHF.I}.

\subsection{Prior}
\label{sec:prior}

We propose using the Berger-M\"uller mixture prior for $\beta_{\ell, m}$
\begin{equation}
	p(\beta_{l,m})  \ = \ \alpha_l \, \delta_0 (\beta_{l, m}) + (1-\alpha_l)\,  \xi_{\tau_l} (\beta_{l,m}),
	\label{eq:HF.prior}
\end{equation}
where $\xi_{\tau}(\beta) \ = \ \tau \xi (\tau \beta)$,  $\delta_0 (x)$ is the Dirac-delta function at zero, $\alpha_l$ is the prior probability that the wavelet coefficient is zero, $\tau_l$ is the prior precision and $\xi$ is the distribution of a non-zero wavelet coefficient. \cite{Johnstone:05} recommended using a heavy-tailed distribution, such as the Laplace distribution, to model this parameter and we use this here. Therefore
\begin{equation}
	p(\beta_{l,m})  \ = \ \alpha_l \, \delta_0 (\beta_{l, m}) + \tfrac{1}{2}(1-\alpha_l) \, \tau_l \, \exp\leftb\{ - \tau_l \, |\beta_{l,m}|\right\},
	\label{eq:L.prior}
\end{equation}
where $\tau_l$ is the prior precision and $2\tau_l^{-2}$ is the prior variance for scale $l = 1, \ldots, J$.

\subsection{Hyperparameter Determination}
As in \citet{Johnstone:05} we use marginal maximum likelihood estimation (MMLE) to determine the hyperparameters: prior probability and precision ($\alpha_l, \, \tau_l$), and error variance $\nu_l$. To do this, we maximize the hyperparameters over the log-likelihood of the error distribution multiplied by the prior,
\begin{equation}
	\mathcal{L}(\alpha_l, \tau_l, \nu_l,| \mathbf{h}_l) \ = \ \sum_{m=0}^{2^l-1} \ \log \leftb\{\alpha_l\phi_{\nu_l} (h_{l,m}) + (1- \alpha_l) \gamma(h_{l,m}| \nu_l, \tau_l) \right\},
	\label{eq:HF.LL}
\end{equation}
where
\begin{equation}
	\gamma(y| \nu_l, \tau_l) \ = \ \int_{-\infty}^\infty \phi_{\nu_l}(y-x)  \, \xi_{\tau_l} ( x) dx.
	\label{eq:HF.gamma}
\end{equation}
The maximum log-likelihood can not be obtained analytically and required numerical maximisation.

\subsection{Likelihood}
\label{sec:likelihood}

Due to Property~6.1~(4) the Haar-Fisz transformation bestows approximate/asymptotic Gaussianity upon the data. Hence, we assume a likelihood of the form
\begin{equation}
	p(h_{l,m}| \beta_{l,m}) \ = \ \phi_{\nu_l} ( \beta_{l,m}-h_{l,m}) \ = \nu_l^{-1} (2\pi)^{-1/2} \exp\leftb\{-\frac{1}{2\nu_l^2} (h_{l,m} - \beta_{l,m})^2\right\},
	\label{eq:HF.likelihood}
\end{equation}
where $\phi_{\nu_l}(\cdot)$ is the the probability density function of the Gaussian distribution with variance $\nu_l^2$, which we shall assume is equal to the error variance.

\subsection{Posterior Distribution}

By combining the prior and the likelihood, we obtain the posterior distribution of the form
\begin{eqnarray}
	p(\beta_{l,m}| h_{l,m})  &=&  \frac{p(\beta_{l,m})p(h_{l,m}|\beta_{l,m})}{\int p(y)p(h_{l,m}|y) dy} \notag\\
	&=&   \frac{[\theta_l \delta_0(\beta_{l,m}) + \xi_{\tau_l}(\beta_{l,m}) ] \phi_{\nu_l}(\beta_{l,m}-h_{l,m})}{ \theta_l \, \phi_{\nu_l}(h_{l,m}) + \int \xi_{\tau_l}(y) \, \phi_{\nu_l}(y - h_{l,m})  dy},
	\label{eq:HF.posterior}
\end{eqnarray}
where  $\theta_l = \alpha_l(1-\alpha_l)^{-1}$ is the odds ratio.

We will use the posterior mean as our `estimator' of the wavelet coefficients $\{ \beta_{l, m} \}$. The posterior mean can be obtained by evaluating the integral
\begin{eqnarray}
	\hat{\beta}_{l,m} \  = \ \E[\beta_{l,m}|h_{l,m}]  &=&  \int x \ \frac{p(x)p(h_{l,m}|x)}{\int p(y)p(h_{l,m}|y) dy}  dx\nonumber\\
	&=&  \frac{\int x \, \xi_{\tau_l} (x) \, \phi_{\nu_l}(x - h_{l,m}) dx}{\theta_l \, \phi_{\nu_l}(h_{l,m}) + \int \xi_{\tau_l} (y) \, \phi_{\nu_l}(y-h_{l,m}) dy}.\label{eq:postmean}
\end{eqnarray}

For  confidence intervals we require  the posterior variance which can be calculated via the integral
\begin{eqnarray}
	\var[\beta_{l,m}|h_{l,m}]  &=&  \E[\beta_{l,m}^2|h_{l,m}] - \left(\E[\beta_{l,m}|h_{l,m}]\right)^2\nonumber\\
	&=& \frac{\int x^2 \ \xi_{\tau_l} (x) \, \phi_{\nu_l}(x - h_{l,m}) dx}{\theta_l \, \phi_{\nu_l}(h_{l,m}) +  \int \xi_{\tau_l} (y) \, \phi_{\nu_l}(y-h_{l,m}) dy} \ - \ \hat{\beta}_{l,m}^2.\label{eq:postvar}
\end{eqnarray}
To simplify notation define
\begin{equation}
	Q^i(h_{l,m}) \ = \ \int_{-\infty}^\infty x^i \ \xi_{\tau_l} (x) \, \phi_{\nu_l}(x - h_{l,m}) dx, \qquad \qquad \text{for} \ i = 0, 1, 2.
	\label{eq:Qi}
\end{equation}

\begin{lemma}
	The quantities $Q^i (h)$ for the Laplace mixture prior in \eqref{eq:L.prior} are given by
	\begin{enumerate}[(a)]
		\item $i = 0$
			\[
				Q_l^0(h) \, = \, \frac{\tau_l}{2} e^{-h^2\!/2\nu_l^2}  \left[ e^{\mu_1^2/2\nu_l^2} \Phi\leftb(\frac{- \mu_1}{\nu_l}\right) \ + \ e^{\mu_2^2/2\nu_l^2} \Phi \leftb( \frac{\mu_2}{\nu_l}\right)\right].
			\]
		
		\item $i = 1$
			\begin{align*}
				Q_l^1(h) \, = \, \frac{\tau_l}{2} e^{-h^2\!/2\nu_l^2} & \left\{e^{\mu_1^2/2\nu_l^2} \left[\mu_1 \, \Phi\leftb(-\frac{\mu_1}{\nu_l}\right) - \nu_l \, \phi\leftb(-\frac{\mu_1}{\nu_l}\right) \right]  + \, e^{\mu_2^2/2\nu_l^2} \left[\mu_2 \, \Phi\leftb(\frac{\mu_2}{\nu_l}\right) + \nu_l \, \phi\leftb(\frac{\mu_2}{\nu_l}\right)\right]\right\}.
			\end{align*}
			
		\item $i=2$
			\begin{align*}
				Q_l^2(h) \, = \, \frac{\tau_l}{2} e^{-h^2\!/2\nu_l^2} & \left\{e^{\mu_1^2/2\nu_l^2} \left[(\nu_l^2 + \mu_1^2) \Phi\leftb(-\frac{\mu_1}{\nu_1}\right) - \mu_1 \nu_l \, \phi\leftb(-\frac{\mu_1}{\nu_1}\right) \right] \right. \\
				& \left. \qquad \qquad \qquad \qquad \qquad \quad + \ e^{\mu_2^2/2\nu_l^2} \left[(\nu_l^2 +\mu_2^2) \Phi\leftb(\frac{\mu_2}{\nu_l}\right) + \mu_2 \nu_l \, \phi\leftb(\frac{\mu_2}{\nu_l}\right)\right]\right\}.
			\end{align*}
	\end{enumerate}
	\label{lem:HF.QL012}
\end{lemma}

{\em Proof.} in the appendix.

\begin{proposition}
The posterior mean of the wavelet coefficients in model~\eqref{eq:DWT.HFI} with components specified by sections~\ref{sec:prior} to~\ref{sec:likelihood} is given by:
 
\begin{equation}
	\hat{\beta}_{l,m}  \ = \ \frac{Q^1(h_{l,m})}{\theta_l \, \phi_{\nu_l}(h_{l,m}) + Q^0(h_{l,m})}, 
	\label{eq:HF.post.E}
\end{equation}
and posterior variance by
\begin{equation}
\var[\beta_{l,m}|h_{l,m}] \ = \ \frac{Q^2(h_{l,m})}{\theta_l \, \phi_{\nu_l}(h_{l,m}) + Q^0(h_{l,m})} - \hat{\beta}_{l,m}^2.
	\label{eq:HF.post.Var}
\end{equation}
\end{proposition}

{\em Proof.}  Substitute the formula~\eqref{eq:Qi} into~\eqref{eq:postmean} and~\eqref{eq:postvar}.

The next result gives us the necessary log-likelihood function of our Bayesian model from \eqref{eq:HF.LL} for the Laplace mixture prior.
\begin{lemma}
	The log-likelihood function for the Laplace mixture prior is
	\begin{align*}
		\mathcal{L}&(\alpha_l, \tau_l, \nu_l| \mathbf{h}_l)  \ = \ \sum_{m=0}^{2^l-1} \log\left \{\alpha_l \phi_{\nu_l}(h_{l,m}) +  \frac{\tau_l(1-\alpha_l)}{2} e^{-y^2\!/2\nu_l^2} \left[e^{\mu_3^2/2\nu_l^2} \Phi\leftb(\frac{-\mu_3}{\nu_l}\right) + e^{\mu_4^2/2\nu_l^2} \Phi\leftb(\frac{\mu_4}{\nu_l}\right)\right]\right\},
	\end{align*}
	where $\phi_{\nu}(\cdot)$ is the zero mean Gaussian pdf with variance $\nu^2$, $\Phi(\cdot)$ is the Gaussian cdf, $\mu_3 = y + \nu_l^2 \, \tau_l$ and $\mu_4 = y - \nu_l^2 \, \tau_l$.
	\label{lem:LapLL}
\end{lemma}

{\em Proof.} The result uses the same methods as for the proof of Lemma~\ref{lem:HF.QL012}.

\begin{figure}
	\centering
	\subfiguretopcaptrue
	\subfigure[\normalsize ]{\label{fig:Lap.Post}  \includegraphics[width=0.48\textwidth]{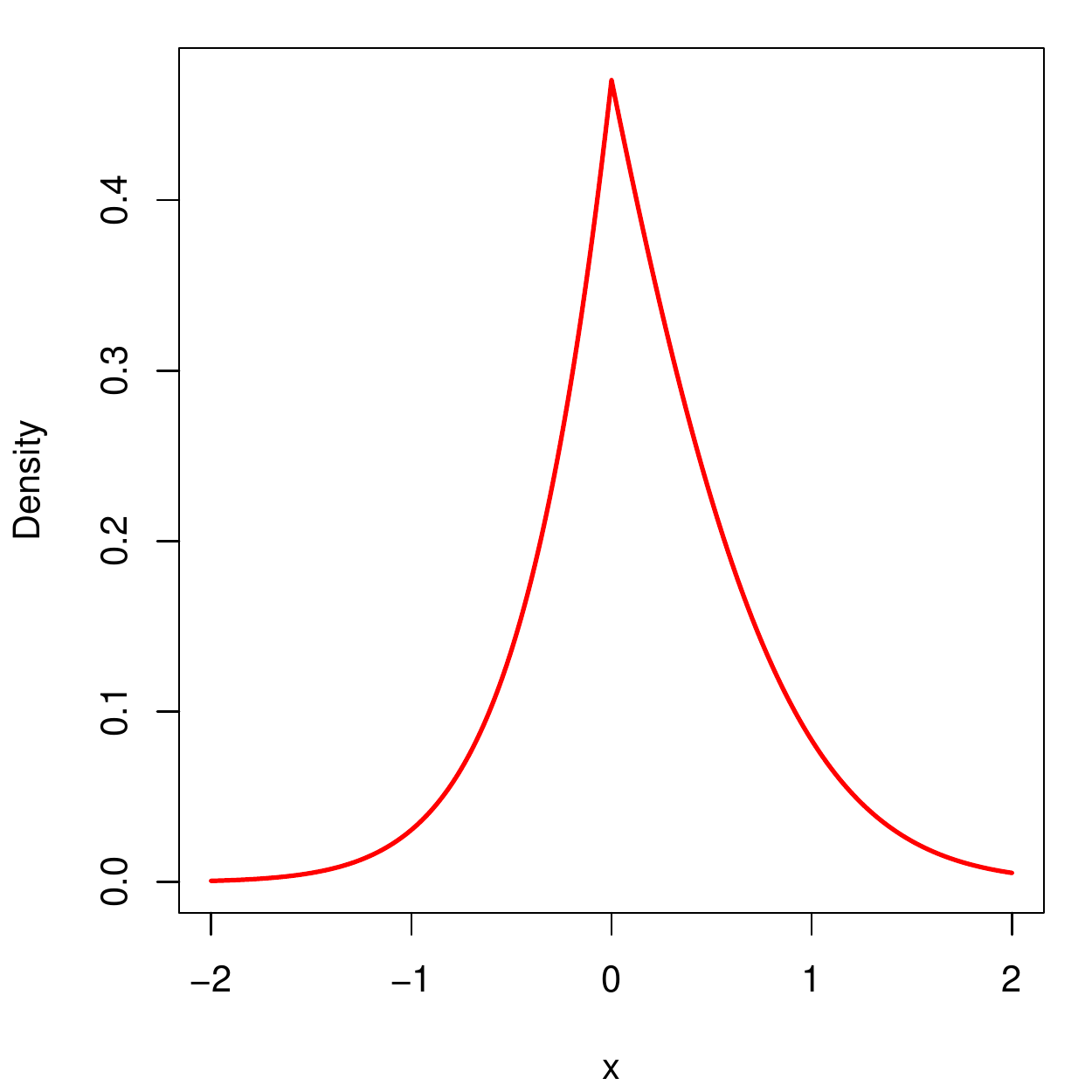}}
	\subfigure[\normalsize ]{\label{fig:Lap.Shrink} \includegraphics[width=0.48\textwidth]{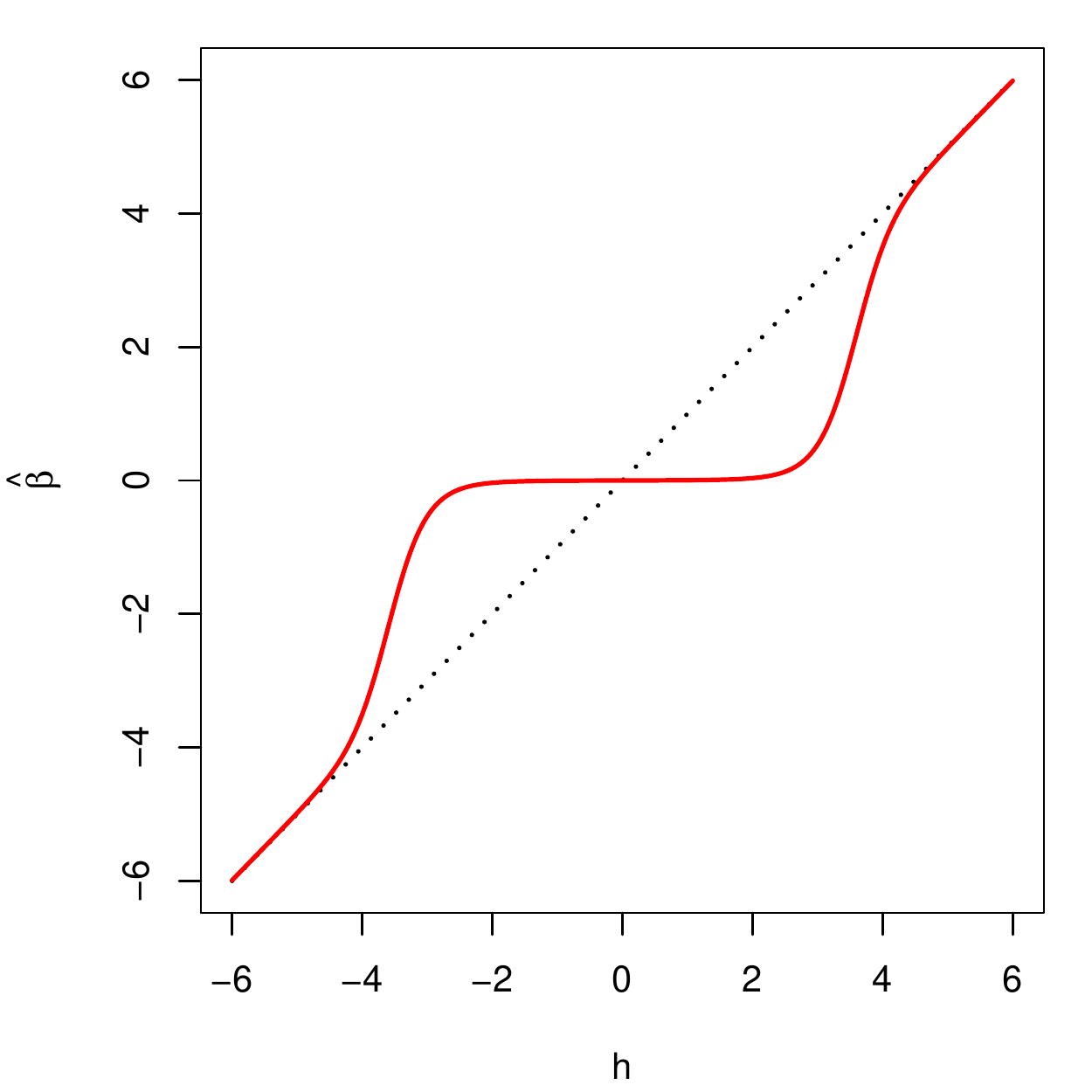}}
	\caption{Plots of the Laplace posterior distribution (a) and shrinkage function (b). Posterior parameters:
	$h=\half, \ \nu=1, \ \theta=\third$ and $\tau = \sqrt{3}$. Shrinkage function for parameters $\nu=1, \ \theta=5$ and $\tau = {}^1\!/_{\!100}$}
\end{figure}

\subsection{Demonstration of Posterior and Shrinkage Function}

Figure~\ref{fig:Lap.Post} shows an example posterior  for the Laplace prior and
figure~\ref{fig:Lap.Shrink} shows
likewise for the shrinkage function.
The latter demonstrates how values of $h$ are `shrunk' to produce the posterior mean estimate, $\hat{\beta}$ with parameters $\nu=1, \ \theta=5$ and $\tau = 1/100$. Values of $|h| \geq 4$ (approximately) remain unchanged, whereas absolute values which are less than four are reduced in magnitude.

\section{Implementation, Simulation and an Example}
\label{sec:Sim}

\subsection{Implementation Issues}

This section describes some of the choices we have made to implement the methods described above in \texttt{R}. 

We determine the hyperparameters via MMLE of \eqref{eq:HF.LL} using the function \texttt{optim} in \texttt{R} which uses the \texttt{L-BFGS-B} method from \citet{Byrd:95}. Empirical investigations revealed that with four coarsest scales, $l = 0, 1, 2, 3$, as they consist of $1, 2, 4$ and $8$ wavelet coefficients (respectively), numerically maximising the log-likelihood for each scale resulted in strongly biased hyperparameter estimates. Therefore, instead of maximising the log-likelihood for the four coarsest scales separately, the coefficients were grouped together and maximisation was performed over all the four scales. To distinguish between scales, the hyperparameter estimates were scaled appropriately, such that as the scale decreased $\alpha_l$ decreased and $\tau_l$ increased by a factor of two.

\begin{figure}
	\centering
	\includegraphics[width=0.8\textwidth]{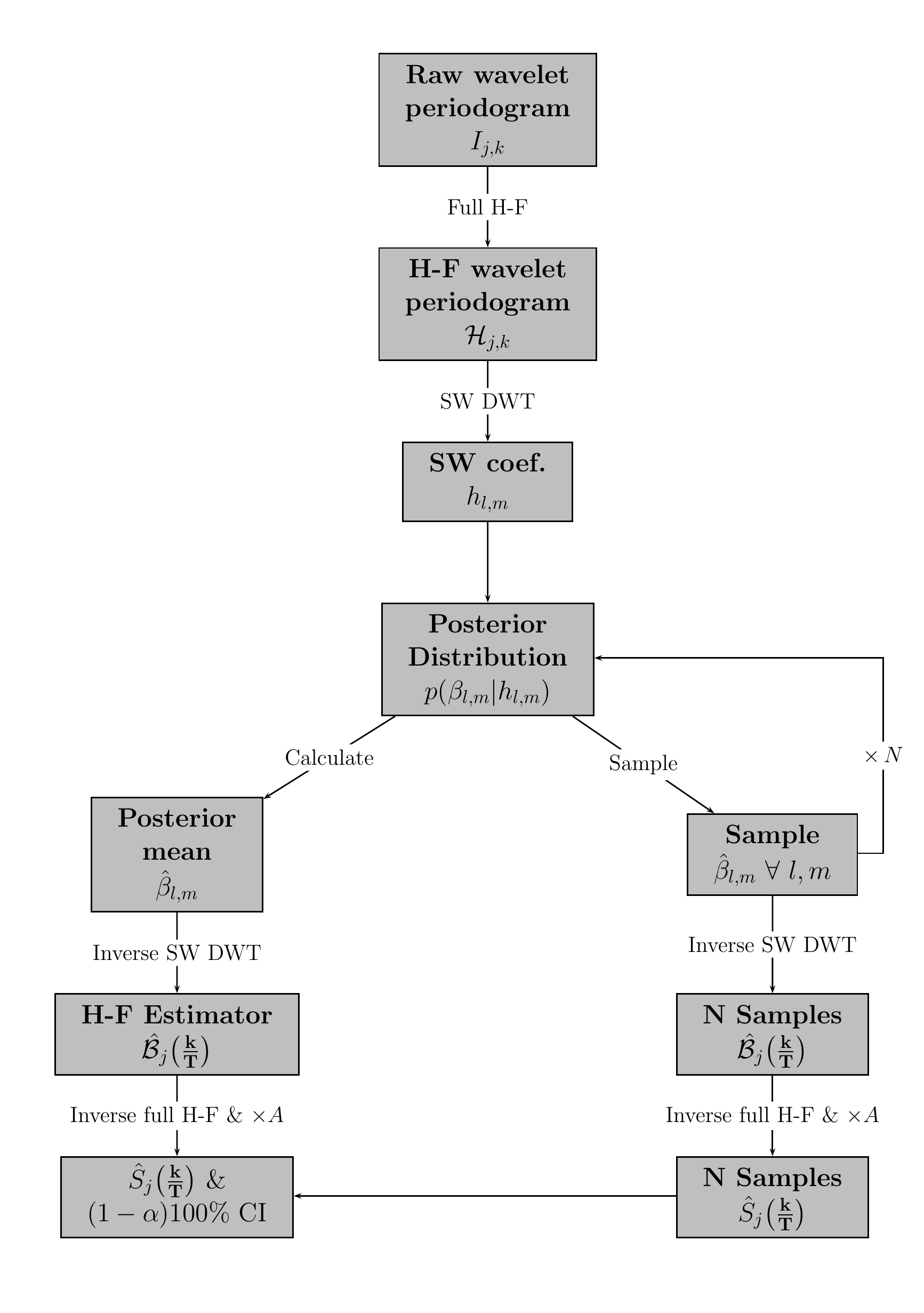}
	\caption{Flow diagram of Bayesian modelling of the discrete wavelet transformation (DWT) of the Haar-Fisz (H-F) transformation of the raw wavelet periodogram using a pre-determined smoothing wavelet (SW).}
	\label{fig:BHF}
\end{figure}

Ultimately, we are seeking an estimate of the posterior (mean and) variance of $\HFB(z)$. Formula~\eqref{eq:HF.post.Var} gives us an estimate of the posterior variance of $\beta_{l, m}$ the wavelet coefficients of $\HFB$. We could use the approximate method of \cite{Barber:01} to obtain the posterior variance of $\HFB(z)$. This works well for Haar wavelets (where the square of the wavelet $\psi^2(z)$ is equal to the father wavelet) but less accurate for non-Haar wavelets. Hence, we adopt the following simple sampling strategy to obtain posterior credible intervals for $\HFB(z)$.

We simulate $S$ realisations for a complete set of wavelet coefficients $\{ \beta_{l, m} \}$ from the posterior distributions given by~\eqref{eq:HF.posterior}. Each realisation of wavelet coefficients is then subjected to the inverse wavelet transform which provides a posterior realisation of the $\mathbf{\HFB} = \{ \HFB(z_1), \ldots, \HFB(z_n) \}$. We then use the sample mean and variance of the $\HFB(z_i)$ to provide the `estimate' and credible intervals.

Figure \ref{fig:BHF} depicts a flow diagram of the entire computational process required to produce an estimate of the EWS via Bayesian wavelet shrinkage of the Haar-Fisz transformed wavelet periodogram and confidence intervals.

\subsection{Simulation}

\begin{sidewaysfigure}
	\centering
	\subfiguretopcaptrue
	\subfigure[\normalsize EWS]{\label{fig:S} \includegraphics[width=0.38\textwidth]{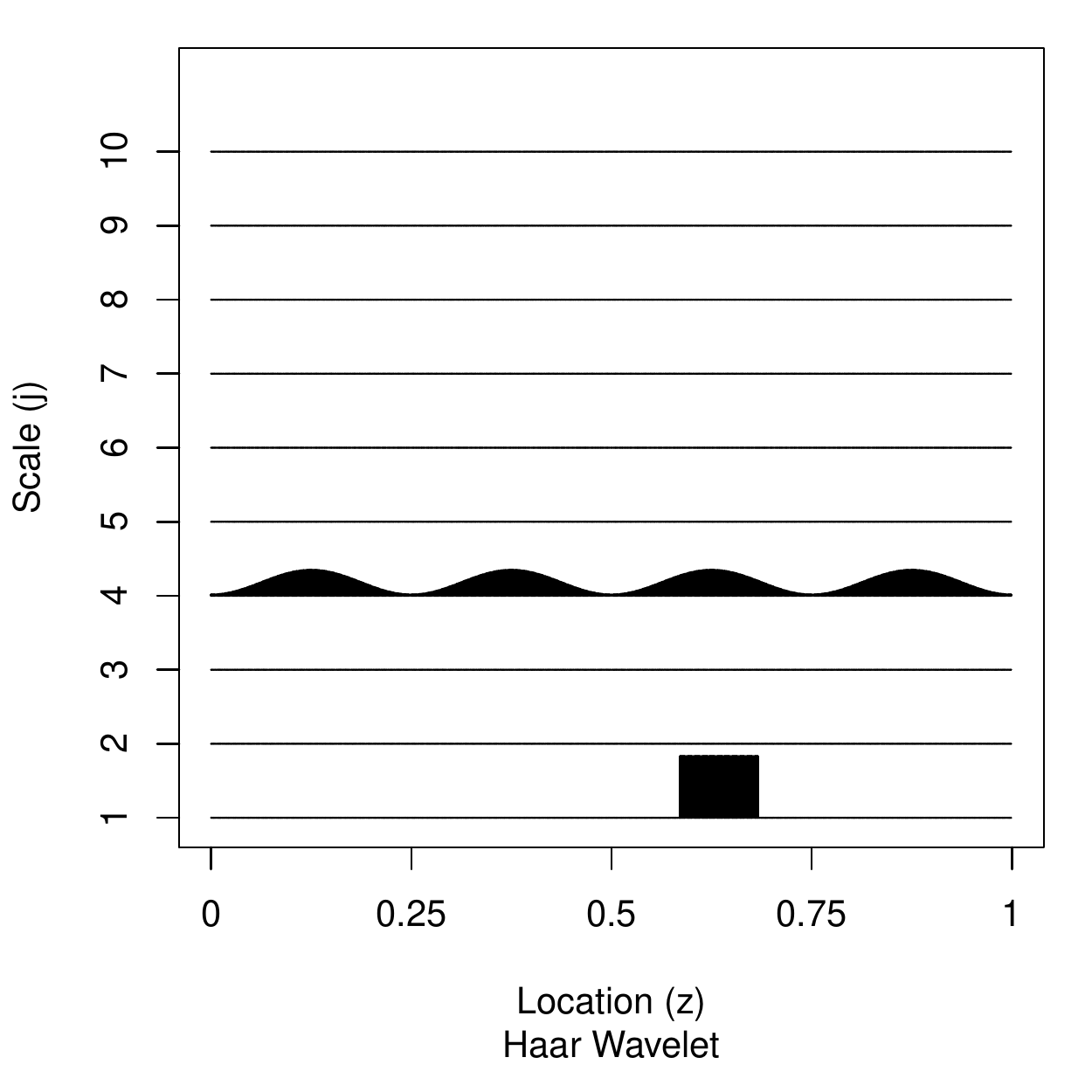}}
	\subfigure[\normalsize Simulated Time Series ($X_t$)]{\label{fig:X.t}\includegraphics[width=0.6\textwidth]{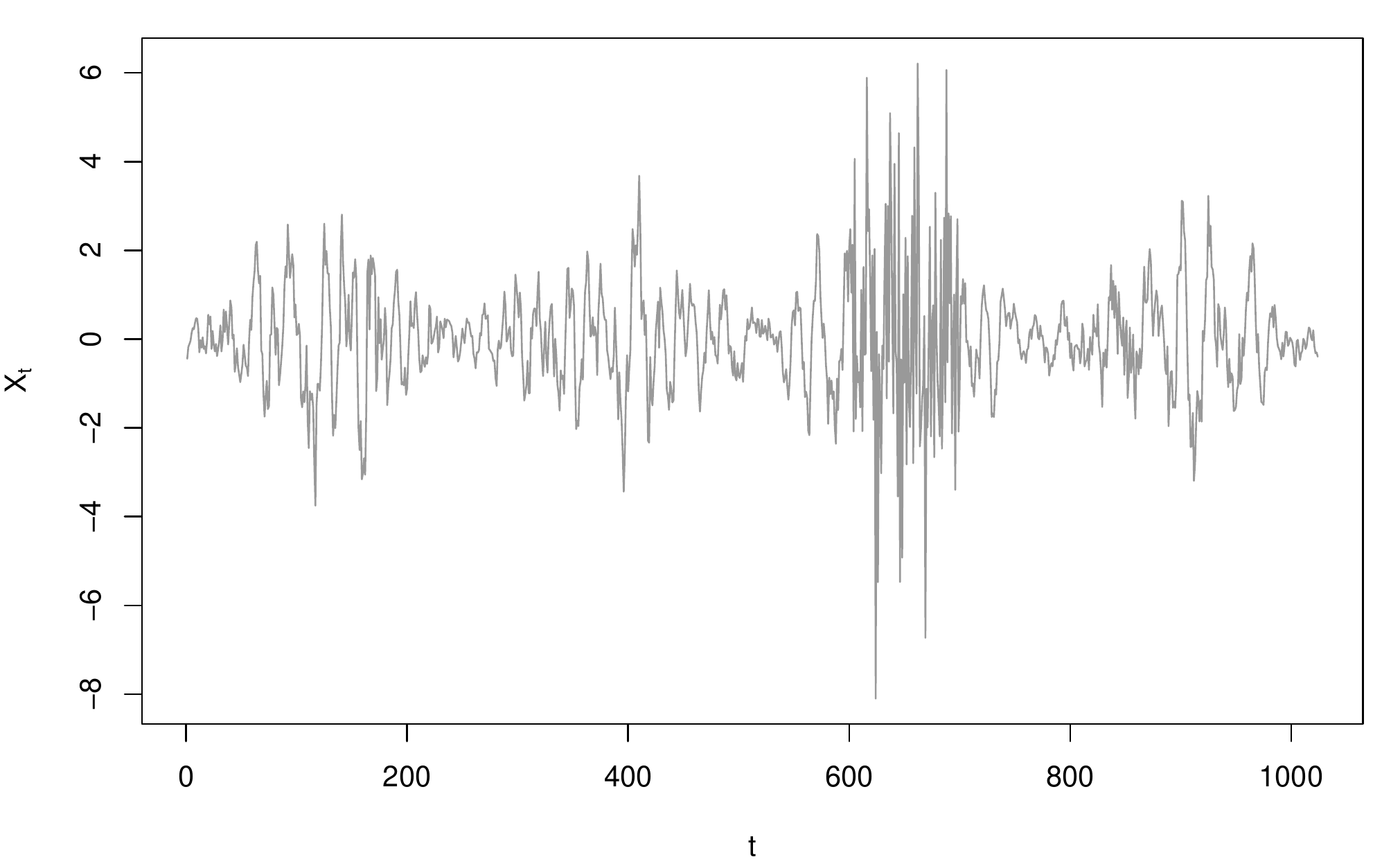}}
	\caption{Plots of the true EWS and a realised LSW process $\{X_t\}_{t=0}^{1023}$ generated using the Haar synthesis wavelet and Gaussian innovations.}
\end{sidewaysfigure}

To test the performance of our method we simulated $200$ realisations, $\{X_t\}_{t=0}^{1023}$, from the EWS in figure \ref{fig:S} with Gaussian innovations as shown in \ref{fig:X.t}. The EWS was designed to encapsulate a time series with slowly varying power at a middle scale along with a burst of power at the finest scale. For each process we produced a Bayesian Haar-Fisz and TI de-noised estimator using the Daubechies extremal phase (EP) with $1-10$ vanishing moments, and Daubechies least asymmetric (LA) with $4-10$ vanishing moments smoothing wavelets. The average mean squared error (AMSE) were calculated using Haar-Fisz estimator with twenty cycle spins to remove any features of the wavelet alignment which might unduly influence our estimator. See \cite{Coifman1} for further details on cycle spinning. 

We calculated the mean EP smoothing wavelet estimate for each of the $200$ processes, then calculated AMSE for both methods. The AMSE for the TI De-noising estimators was $0.186$ and for Bayesian Haar-Fisz estimators $0.127$.

\begin{table}
	\centering
	\begin{tabular}{|l|rr|rr|}
		\hline
		\bf{Vanishing}		& \multicolumn{2}{c|}{\bf{Extremal Phase}} 	& \multicolumn{2}{c|}{\bf{Least Asymmetric}}  \\
		\bf{Moments}		& TI-D		& H-F  		& TI-D		& H-F \\
		\hline
		$1$			& $196$		& $146$ 		& -			& - \\
		$2$			& $198$		& $130$ 		& -			& - \\
		$3$			& $193$		& $123$ 		& -			& - \\	
		$4$			& $191$		& $129$ 		& $196$		& $124$ \\
		$5$			& $190$		& $136$ 		& $195$		& $128$ \\
		$6$			& $188$		& $132$ 		& $195$		& $122$ \\
		$7$			& $187$		& $126$ 		& $196$		& $136$ \\
		$8$			& $186$		& $129$ 		& $195$		& $123$ \\
		$9$			& $186$		& $138$ 		& $195$		& $136$ \\
		$10$		& $185$		& $139$ 		& $195$		& $123$ \\
		\hline
	\end{tabular}
	\caption{Table of average mean square error ($\times 10^{-3}$) over $200$ simulations for the translation-invariant de-noising (TI-D) and Bayesian Haar-Fisz (H-F) estimators using the smoothing wavelets: Daubechies extremal phase (EP) with $1-10$ vanishing moments and Daubechies least asymmetric (LA) with $4-10$ vanishing moments.}
	\label{tab:AMSE.w}
\end{table}

Table \ref{tab:AMSE.w} shows the AMSE for each estimator and choice of smoothing wavelet. The $EP_1$ corresponds to the Haar wavelet, which gives the poorest estimator in both cases, this is only the best wavelet to use if the underlying structure of the EWS for each scale is piecewise constant. We found that both methods seemed fairly robust to the choice of wavelet, as the difference between the AMSE appeared to be fairly small. Although we noticed the AMSE of the TI de-noising estimator decreased as the support of the wavelet increased, which was not the case for the Bayesian Haar-Fisz estimator. However, the Bayesian Haar-Fisz estimator consistently out performed the TI de-noising estimator. 

\begin{figure}
	\centering
	\subfiguretopcaptrue
	\subfigure[\normalsize TI de-noising EWS estimator]{\label{fig:TI.S} \includegraphics[width=0.48\textwidth]{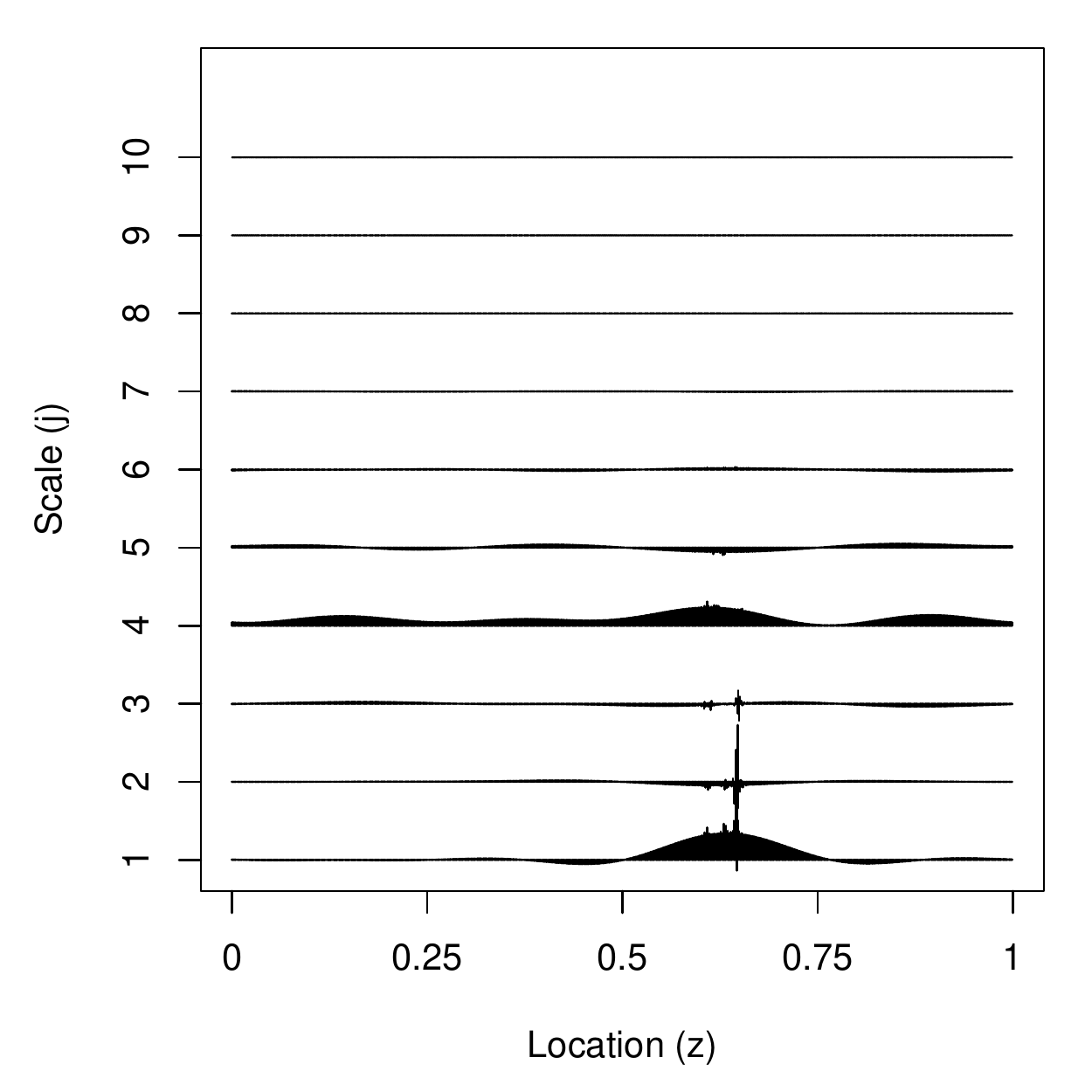}}
	\subfigure[\normalsize Bayesian Haar-Fisz EWS estimator]{\label{fig:HF.S} \includegraphics[width=0.48\textwidth]{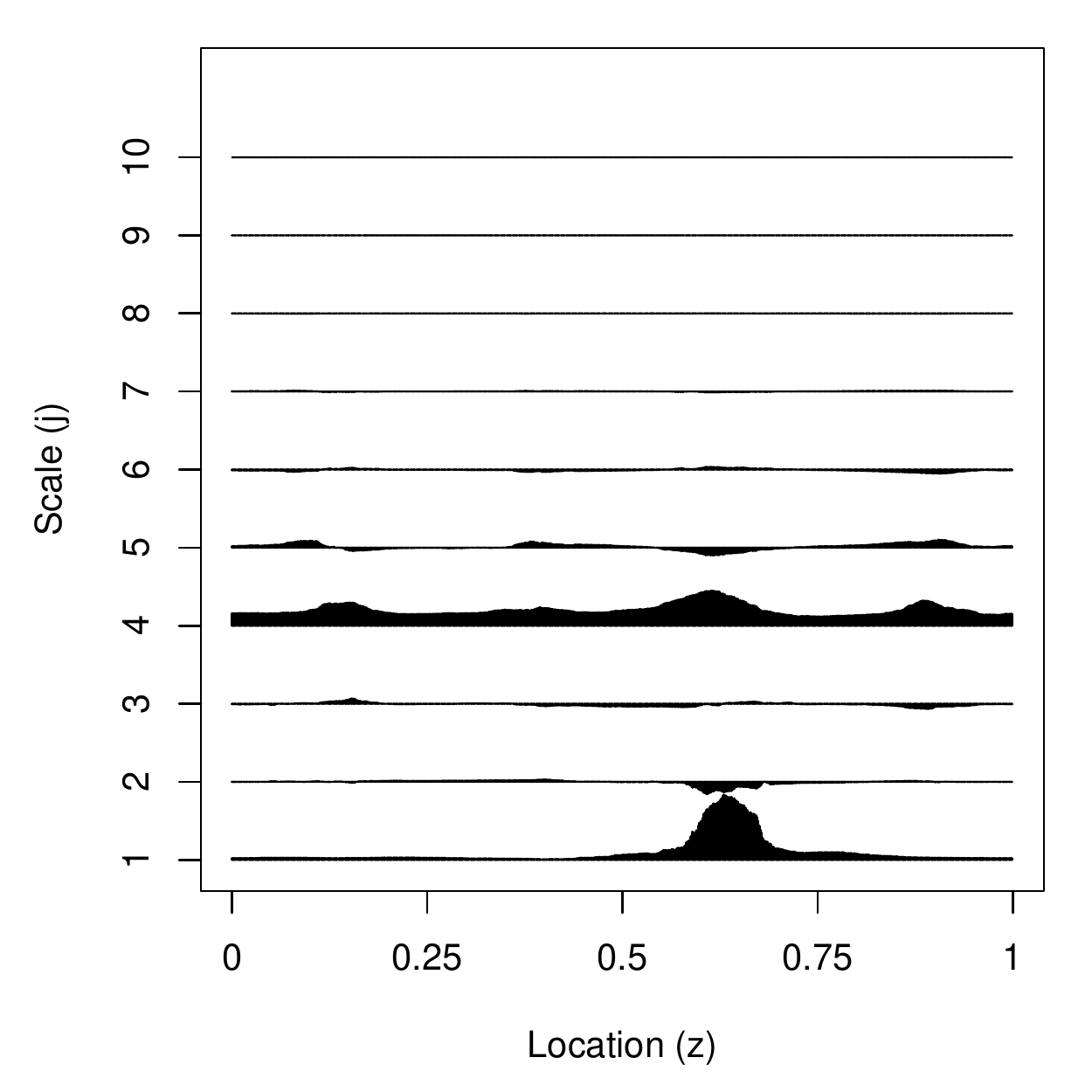}}
	\caption{Plots of the estimated EWS using TI de-noising with SW = $EP_{10}$ and Bayesian Haar-Fisz estimators with SW = $LA_6$, arising from a realisation from the simulated spectrum show in Figure~\ref{fig:S}.}
\end{figure}

We compared our best estimator using Bayesian modelling of the Haar-Fisz periodogram (SW = $LA_6$), see figure \ref{fig:HF.S}, with the best TI de-noising estimator \citep[SW = $EP_{10}$]{Nason:00}, as shown in figure \ref{fig:TI.S}, determined from the results in table \ref{tab:AMSE.w}.

Comparing the plots in figures \ref{fig:HF.S} and \ref{fig:TI.S}, we can see that the Bayesian Haar-Fisz estimator is less susceptible to Gibbs-type phenomena, but the leakage of power in neighbouring scales appeared to be fairly comparable for both estimators. Some of the power from scale $j=6$ has leaked into $j=5,7$, which has made recovery of the true underlying signal difficult.

\begin{figure}
	\centering
	\subfiguretopcaptrue
	\subfigure[\normalsize $S_1(z)$]{\label{fig:S1} \includegraphics[width=0.48\textwidth]{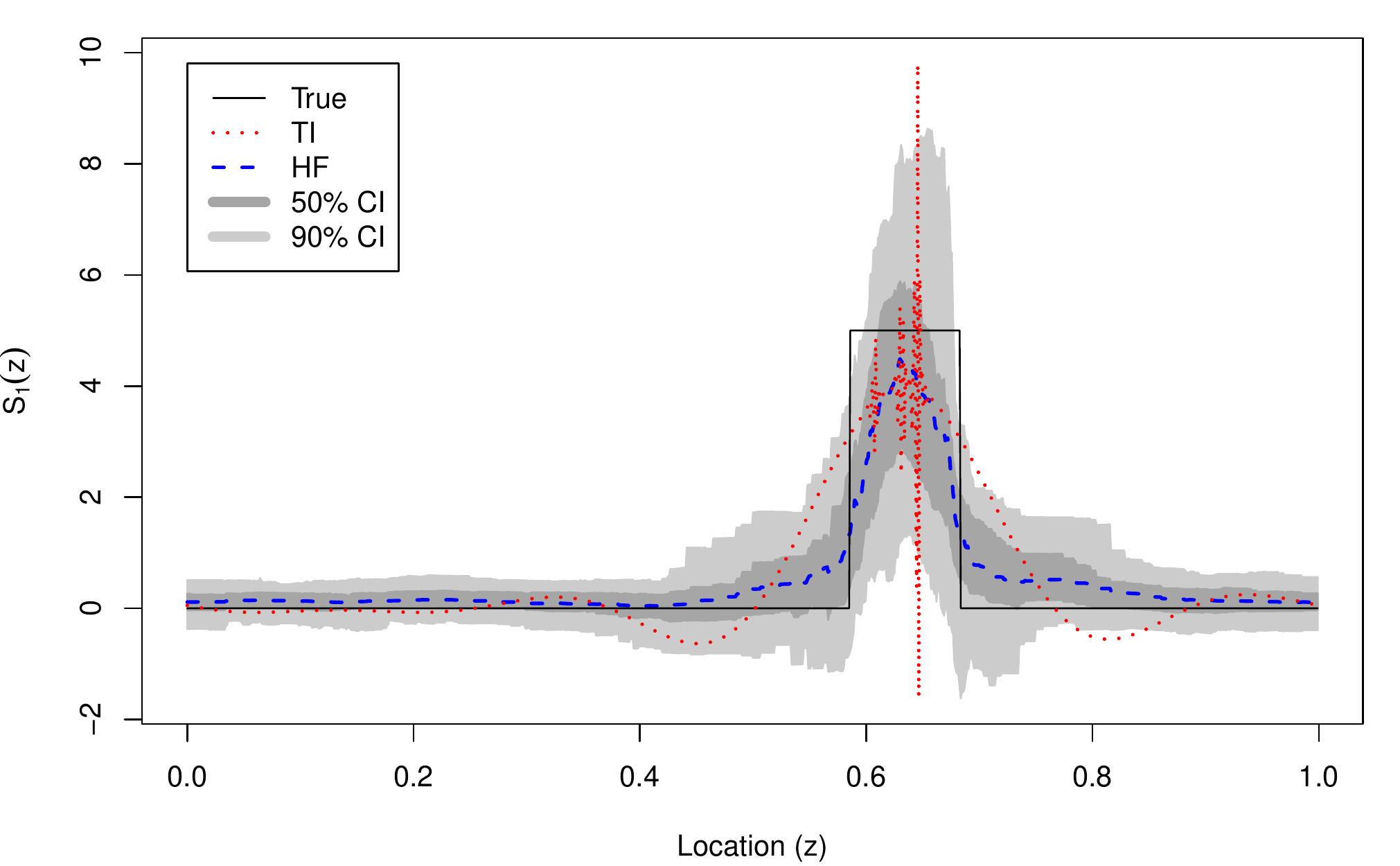}} 
	\subfigure[\normalsize $S_2(z)$]{\label{fig:S2} \includegraphics[width=0.48\textwidth]{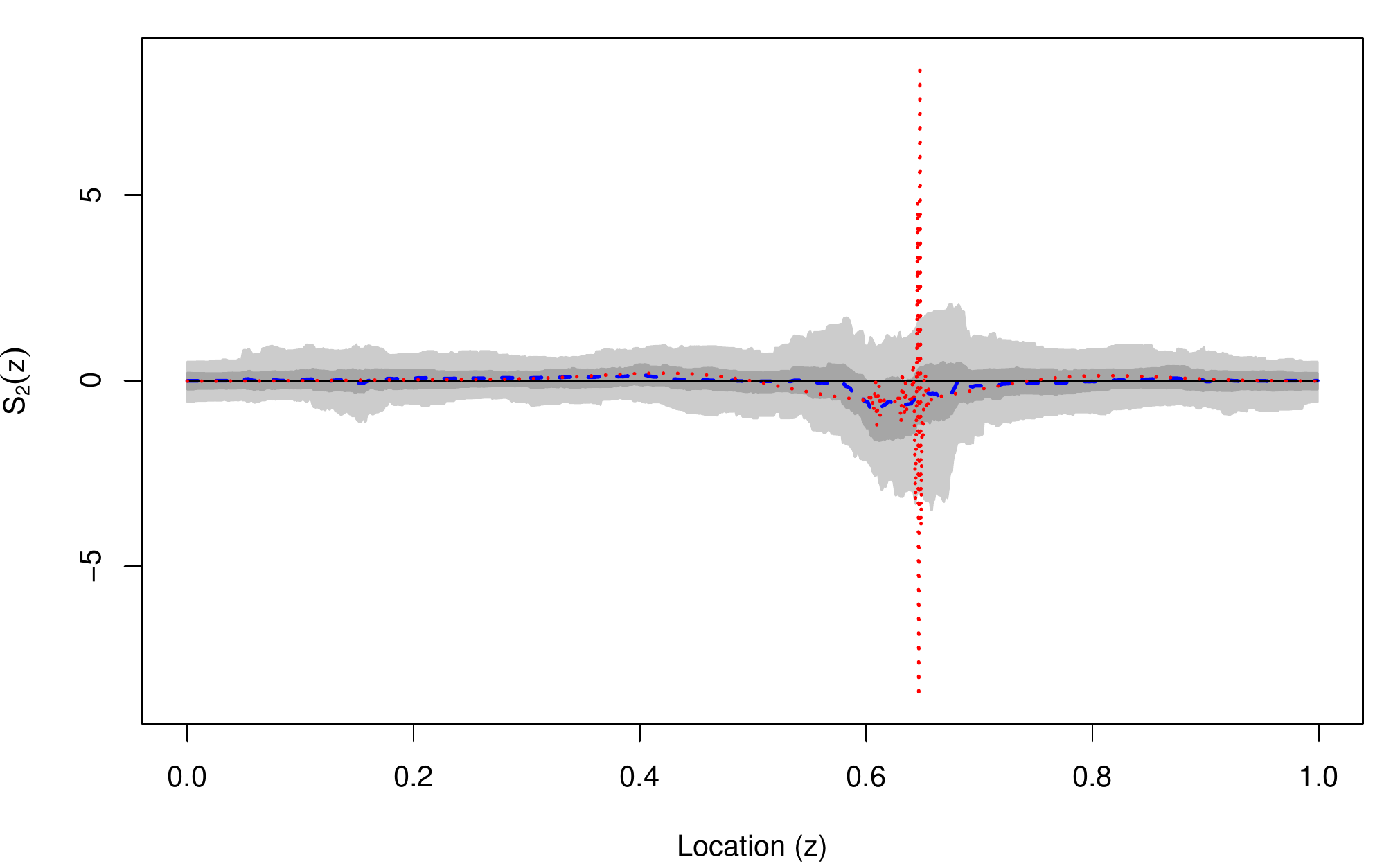}}\\
	\subfigure[\normalsize $S_3(z)$]{\label{fig:S3} \includegraphics[width=0.48\textwidth]{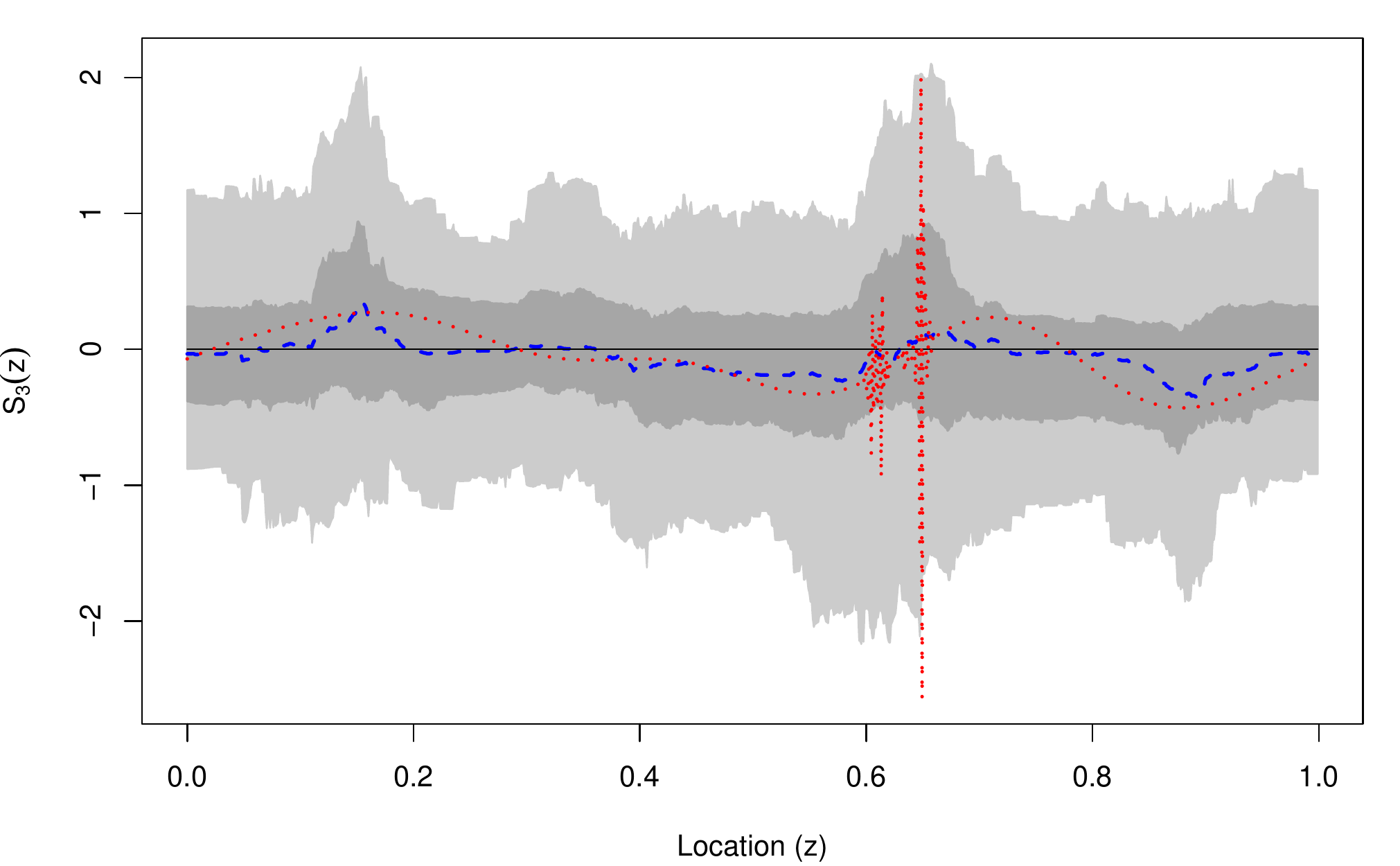}}
	\subfigure[\normalsize $S_4(z)$]{\label{fig:S4} \includegraphics[width=0.48\textwidth]{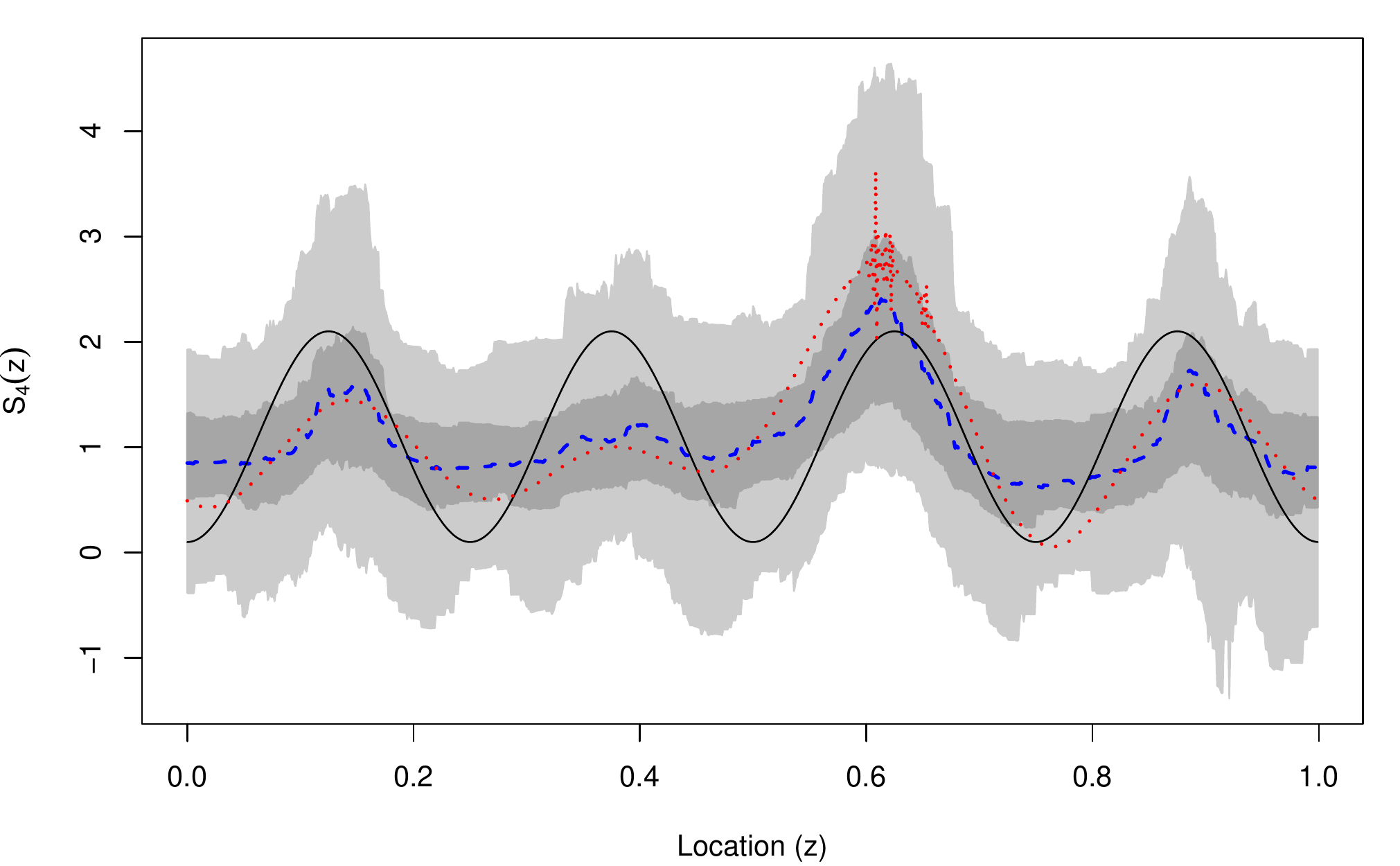}}\\
	\caption{Plots of the true EWS (black solid line) TI de-noising estimator (red dotted line) and Bayesian Haar-Fisz estimator (blue dashed line) for the first ($j=1$), second ($j=2$), third ($j=7$) and fourth ($j=6$) finest scales, with the 50\% (dark grey) and 90\% (light grey) confidence intervals for the Bayesian Haar-Fisz estimator.
	These estimates are all obtained from a single realisation from the
	spectrum shown in Figure~\ref{fig:S}.}
\end{figure}

Figure~\ref{fig:S4}--\ref{fig:S1} show the EWS estimation for the simulated example in greater detail. The new method is certainly better at detecting the burst at the finest scale shown in Figure~\ref{fig:S1}. In Figure~\ref{fig:S4} we judge our method to be comparable to the TI-denoising away from $z=0.6$ and considerably better near to $z=0.6$.

A key advantage of our new methodology is the ability to easily generate credible intervals which are shown by grey-scale in Figures~\ref{fig:S4}--\ref{fig:S1}.
For example, even though the estimator for $S_3(z)$ appears to be non-zero in figure \ref{fig:S3}, the 50\% credible intervals completely contain zero which indicates (correctly) that there is no real power at this scale level. The same is true, but less clear maybe, in Figure~\ref{fig:S2}.

\subsection{ECG Example}

To test our methods further, we consider the study of infant sleep \citep{Sawczenko:95}. Five mothers and their healthy first-born infants slept in a sleep laboratory designed to be similar to a normal domestic bedroom once a month for the first five months. The rooms were thermally controlled  and all infants slept supine in a cot besides their mother, who were free to care for their infants as they would at home (e.g.\ feed, change nappy, etc). Most studies commenced around 8-9pm and finished around 8-9am the next morning. 

Amongst the measurements taken of each infant was their heart rate via ECG (electro-cardiogram) monitors, their brain waves via a EEG (electro-encephalogram) sensor and eye movements using a EOG (electro-oculogram) sensor. The infant's sleep state was then determined through manual analysis where a trained observer visually interprets the EEG and EOG at predetermined time periods, which can be time consuming and laborious. Four sleep states were recorded: AWAKE, ACTIVE SLEEP, BETWEEN and QUIET SLEEP. For simplicity, we have combined the latter three states into ASLEEP.

\begin{figure}
	\centering
	\includegraphics[width=0.95\textwidth]{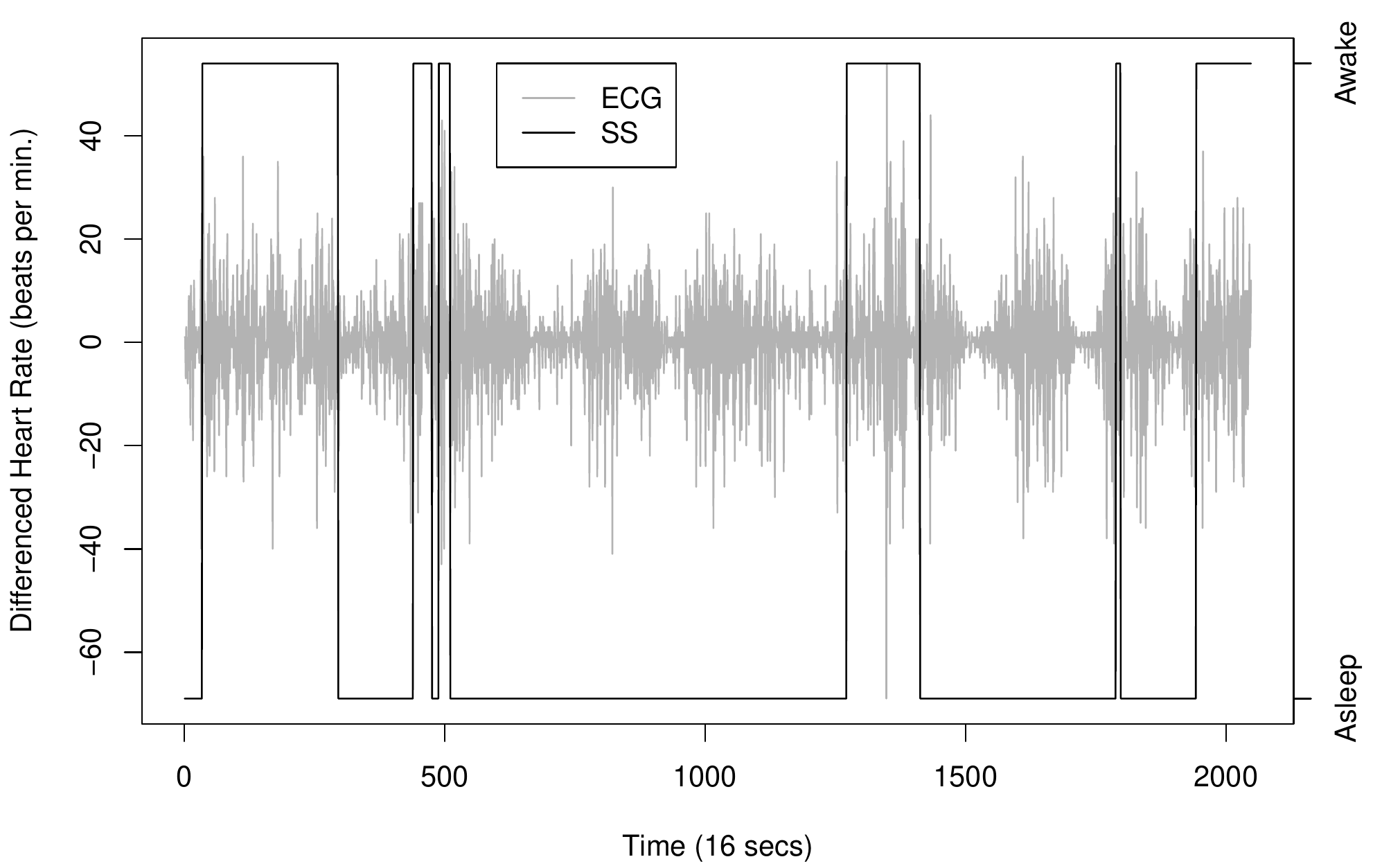}
	\caption{A plot of the ECG (light grey line) and sleep state (black solid line) of a 66 day old infant sampled every 16 seconds recorded from 21:17:59 to 06:27:18.}
	\label{fig:Baby}
\end{figure}

Figure \ref{fig:Baby} is a plot of 2048 observations sampled every 16 seconds recorded from 21:17:59 to 06:27:18 of the ECG and determined sleep state for the same sixty-six day old infant. The plot indicates that when the infant is awake there is a larger variance in the infant's heart rate compared to the two different sleep stages, for which quiet sleep appears to possessing the smallest variance. We have produced an estimate of the EWS for the differenced ECG data to establish whether we could use the second order structure of the data to determine the infant's sleep state.

\begin{sidewaysfigure}
	\centering
	\subfiguretopcaptrue
	\subfigure[\normalsize Estimated Infant ECG EWS]{\label{fig:InfantS} \includegraphics[width=0.38\textwidth]{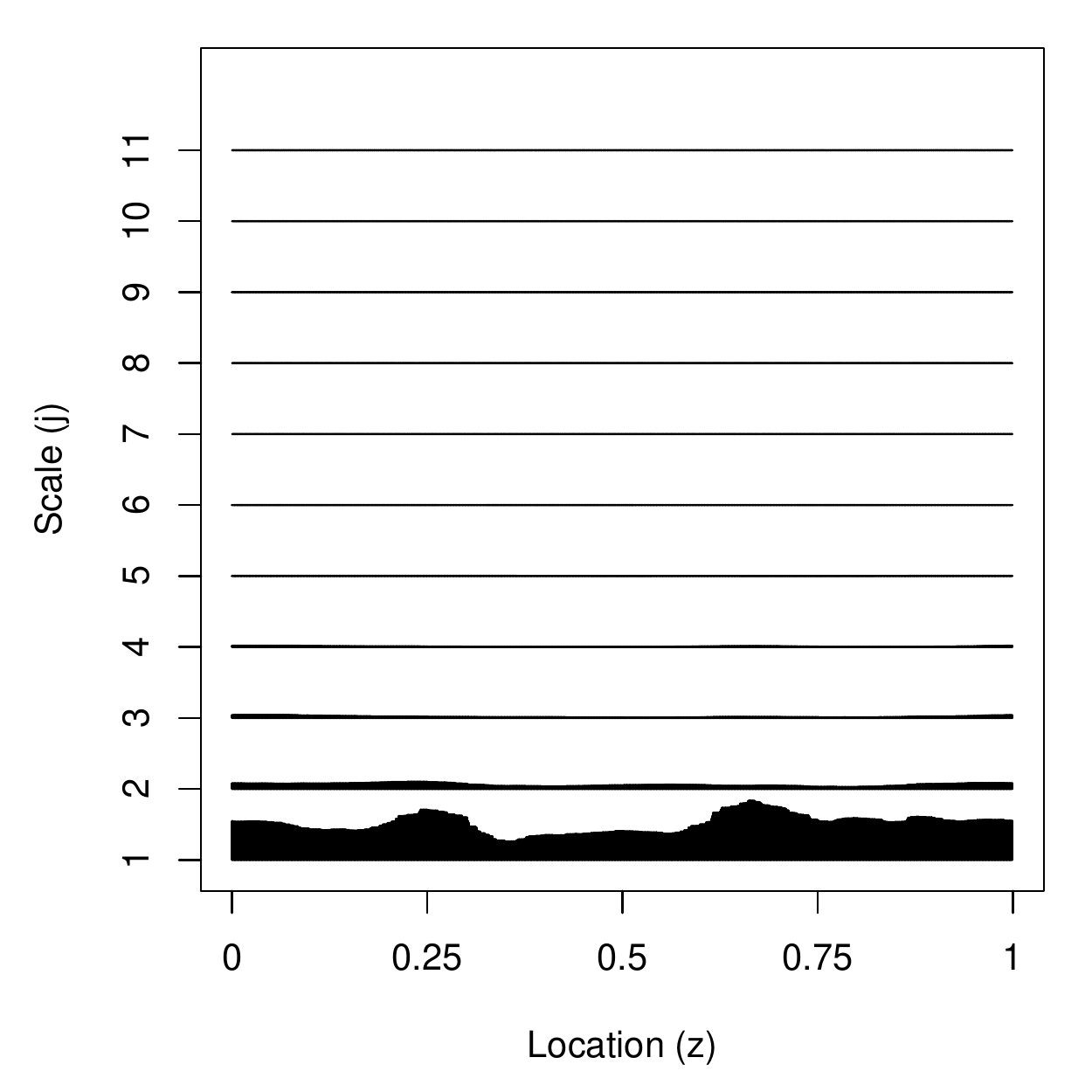}}
	\subfigure[\normalsize Estimated Infant ECG at the finest scale with CI]{\label{fig:InftantS10
}\includegraphics[width=0.6\textwidth]{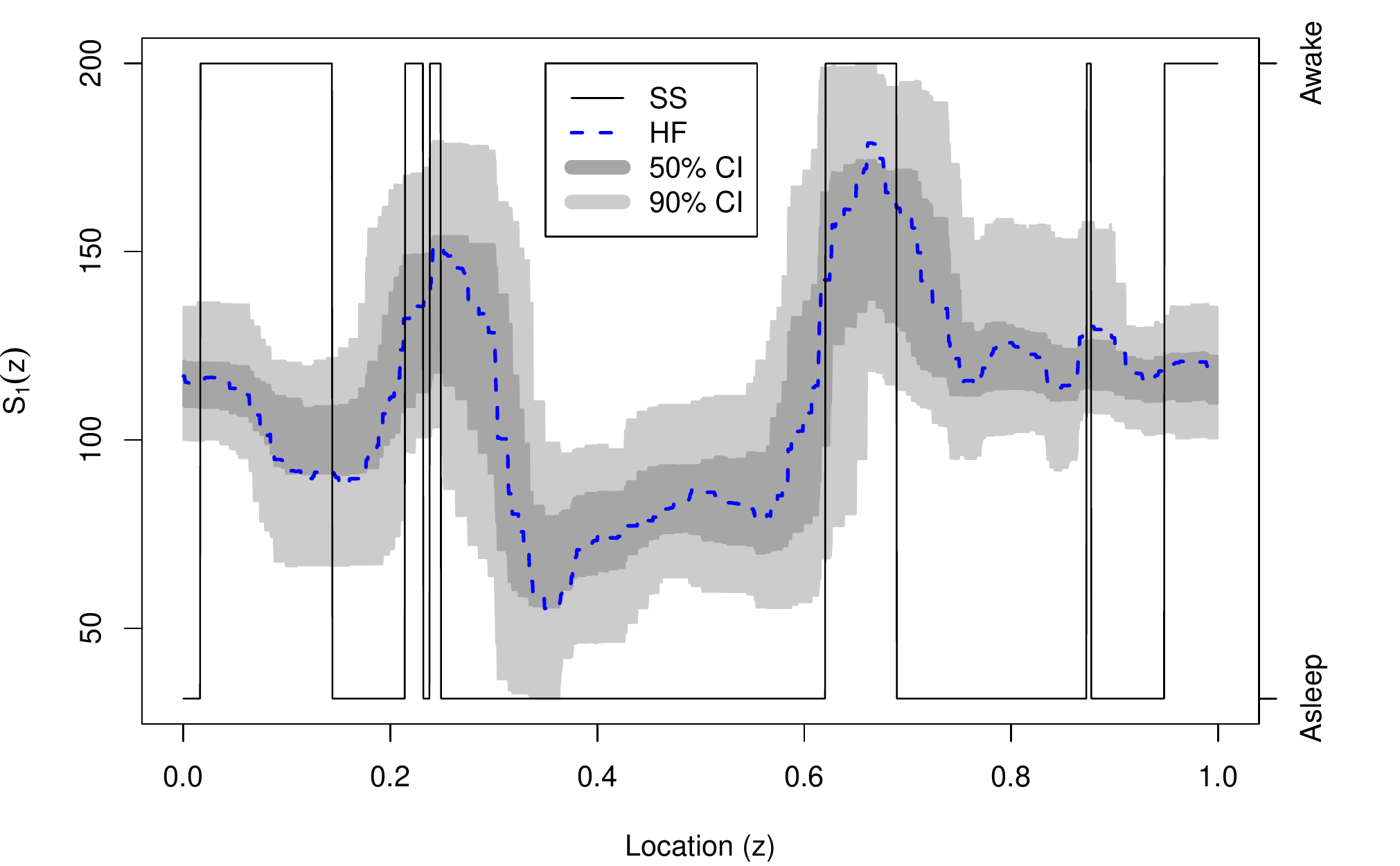}}
	\caption{(a) A Plot of the estimated EWS for all scales. (b) A Plot of the estimated EWS (blue dashed line) for $j=10$ with 50\% (dark grey) and 90\% (light grey) CI for the differenced Infant ECG data and sleep state (black solid line).}
\end{sidewaysfigure}

The plot in \ref{fig:InfantS} implies the majority of the power of the spectrum is present at the finest scale. There appears to be some difficulty in discerning the infant's sleep state when it changes quickly (such as between location $z \in [0.2, 0.4]$). As with earlier analyses, such as that in \cite{Nason2}, there appears to be a link between active sleep and higher power at the finest scale. However, our new analysis reveals much more: that there is more uncertainty associated with the higher power estimates and more certainty when the power is lower. The arrangement of the posterior mean estimate relative to the 50/90\% credible intervals indicates skew in the posterior distribution which is especially noticeable around the peak near to 0.65.

\section{Conclusion and Further Work}
\label{sec:conc}

This article combines the Haar-Fisz transform with Bayesian wavelet shrinkage to obtain a  new method for modelling the evolutionary wavelet spectrum of a locally stationary wavelet process. Bayesian wavelet shrinkage is known and powerful technique and well-established for noisy data contaminated by uncorrelated Gaussian noise which the Haar-Fisz transform approximately, but effectively, provides. Although there are competing methods for spectral estimation there are, as far as we know, no methods for generating confidence intervals for evolving spectra certainly in the wavelet case. Our Bayesian wavelet shrinkage gives a rational method for assessing uncertainty in this case providing us with approximate credible intervals.

Further work to improve our method would be to improve our method of determining hyperparmeters and also investigate its application to irregularly spaced time series. Another interesting possibility is to apply Bayesian wavelet shrinkage to Haar-Fisz transformed spectra in the stationary or locally stationary Fourier case.

\section{Acknowledgements}
KS was supported by a studentship funded by the SuSTaIn Science and Innovation Award grant EP/D063485/1. GPN was partially supported by EPSRC grants from EP/I01687X/1: ``The Energy Programme,  an RCUK cross-council initiative led by EPSRC and contributed to by ESRC, NERC, BBSRC and STFC''.

\appendix
\section{Proofs}

\subsection*{Proof of Lemma~\ref{lem:HF.QL012}.}

The integral in~\eqref{eq:Qi} can be shown to equal to:
\begin{equation}
Q_l^i(h_{l,m})  =  \frac{\tau_l}{2} e^{-h_{l,m}^2/2\nu_l^2} \left[e^{\mu_1^2/2\nu_l^2} \int_{-\infty}^0 y^i \, \phi_{\nu_l}(y - \mu_1) \, \diff y  \ + \  e^{\mu_2^2/2\nu_l^2} \int_{-\infty}^0 (-y)^i \, \phi_{\nu_l}(y + \mu_2) \, \diff y \right],
\label{eq:newQiform}
		\end{equation}
where $\phi_{\nu}(\cdot)$ is the zero mean Gaussian pdf with variance $\nu^2$, $\mu_1 = h_{l,m} + \nu_l^2 \, \tau_l$ and $\mu_2 = h_{l,m} - \nu_l^2 \, \tau_l$. Formula~\eqref{eq:newQiform} is obtained by substituting in the formula for the Laplace density in~\eqref{eq:Qi} and splitting the integral into two parts on the negative and positive domains. Then, on each integral, the $\exp (-\tau_l |x|)$ term is merged with the exponential in the normal density and then the square completed for each term.

Finally, to obtain the quoted formulae in Lemma~\eqref{lem:HF.QL012} use the following properties of the Gaussian distribution:
	\[
		\int_{-\infty}^y \phi (x) \diff x = \Phi(y), \qquad \int_{-\infty}^y x \phi(x) \diff x = -\phi(y)  \qquad \text{and} \qquad  \int_{-\infty}^y x^2 \phi (x) \diff x = \Phi(y)-y\phi (y).
	\]
	\qed

\bibliographystyle{asgm}
\bibliography{Bibliography}

\end{document}